\documentclass[prb,aps,twocolumn,groupedaddress,floats,showpacs,final]{revtex4}
\pdfoutput=1
\usepackage{graphicx}
\usepackage{dcolumn}
\usepackage{bm}
\usepackage{amsmath}
\usepackage{gensymb}
\usepackage{color}
\usepackage{ulem}
\definecolor{blue}{rgb}{0.3,0.3,0.9}
\pdfoutput=1
\def\beq{\begin{eqnarray}}
\def\eeq{\end{eqnarray}}

\begin{document}

\author{Max Yarmolinsky and Anatoly Kuklov}
\address{$^1$ Department of Physics \& Astronomy, CSI, and the Graduate Center of CUNY, New York.}


\title{Strong and weak field criticality of 2D liquid gas transition}

\date{\today}
\begin{abstract}
Finite size scaling (FSS) analysis of the liquid gas criticality is complicated by the absence of any broken symmetry. This, in particular, does not allow a straightforward finding of the coexistence line and the critical point. 
The numerical flowgram (NF) method \cite{Annals} is adapted for a controlled determination of  the coexistence line and the critical point, with the critical indices $\mu,\, \nu$ measured within 1-2\% of the total error. The approach based on the NF for measuring the non-analytical corrections to the {\it diameter} -- the mean density of the liquid and gas phases along the coexistence line -- is outlined. Our analysis is a first step toward a  general evaluation of isolated critical and multicritical points.
\end{abstract}


\maketitle

\section{Introduction}

The liquid-gas (LG) transition in free space is a cornerstone and testbed for the theory of phase transitions. It features a first order transition and an isolated critical point. Most importantly, the critical point is characterized by emerging $Z_2$ symmetry, which is not explicitly present in the Hamiltonian.  There is no actual broken symmetry in either phase, and the role of the order parameter is played by the difference in densities of liquid and gas (see in Ref.~\cite{Landau}). 
Nevertheless, it was suggested that the LG criticality belongs to the Ising universality class (see in Refs. \cite{Landau,Kadanoff,Patashin}) with the linear mixing of the primary scaling operators resulting in  non-analytical corrections to the so called {\it diameter}  \cite{Mermin,Pokrovskii,Patashin,Bruce,Wilding,Fisher2000,Fisher2001,Fisher2003}.  

 The theme of symmetry emergence at the point of a continuous transition permeates current physics. Among early examples beyond the LG criticality is the emergence of the isotropic Heisenberg fixed point in the Hamiltonian with cubic anisotropy \cite{Aharony,Nelson}. Later the conjecture of the SO(5) symmetry enlargement was put forward in connection with high temperature superconductivity \cite{SO5}. A general renormalization group analysis for the possibility of the symmetry enlargement from O(N$_1$)$\times$O(N$_2$) to O(N$_1$+N$_2$) has been conducted in Ref.\cite{Vicari}. More recently, a theory of the so called Deconfined Critical Point (DCP) has been suggested \cite{DCP}. It describes a generic (tentatively) continuous transition between phases characterized by completely different order parameters and an extended emerging symmetry not present in the Hamiltonian.
This theory, however, is plagued by uncertainties in the type of the transition. The numerical flowgram method \cite{Annals} has proved to be effective in resolving the controversy for the effective field description of the DCP and establishing that the transition is actually of weakly first order \cite{Annals,DCP_us}.
[This, however, does not exclude a possibility that a specifically tuned microscopic model \cite{Sandvik_JQ} exhibits a continuous transition  
which is not fully captured by the effective description \cite{DCP}].

In the present work we extend the method \cite{Annals} (further developed in Ref.\cite{NJP}) to the LG criticality -- as a first step toward analyzing other  multicritical points. 

The nature of the LG critical point has attracted a lot of experimental \cite{Wulkowitz,Garland,Hayes,Moses,Moses_2D}, analytical and numerical  \cite{SW, Fisher2000, Fisher2001,Fisher2003, Watanabe} efforts. The FSS analysis \cite{FSS} of the 3D LG systems has been conducted in Refs.\cite{Fisher2003} and strong numerical arguments in favor of the Ising universality have been presented. Despite that some important questions are still not answered -- such as an accurate determination of the coexistence line and the critical point. Accordingly, the  
 non-analytical corrections to the diameter \cite{Mermin,Pokrovskii,Patashin,Bruce,Wilding,Fisher2000,Fisher2001,Fisher2003} remain to be detected.
It is important to mention that the standard numerical approach to the LG criticality relies on finding the coexistence line by the positions of the peaks of  the density histogram and their further extrapolation toward the critical point (with its position also unknown and treated as a fitting parameter). 
Close to the critical point the impact of the fitting errors  of the critical temperature $T_c$ and pressure $P_c$ (or density) on the critical indices are progressively increasing. 
Thus, the final uncertainty in the critical indices becomes large and virtually uncontrolled.  

It should also be mentioned that the experimental results are not without controversy -- while some experiments \cite{Wulkowitz,Garland} claimed non-Ising universality, others \cite{Moses,Moses_2D} favored the conjecture of the emergent $Z_2$ symmetry.   Despite the fact that the $Z_2$ nature of the LG criticality is commonly accepted, the accuracy of the measured exponents (experimental and numerical) leaves some room for doubt \cite{Bondarev}. To great extent this is also due to relying on multi-parametric fits using the position of the critical point as a fitting parameter.  

Very recently, we have applied the NF method \cite{Annals,NJP} to the LG criticality in 2D within the Grand Canonical Ensemble Monte Carlo simulations\cite{Max}. The main advantage of this approach is that the position of the critical point in the plane of temperature, $T$, and chemical potential, $\tilde{\mu}$, is obtained as a byproduct of tuning the system into the critical range of a Binder cumulant  $U_B$ \cite{Binder}. 
It is important to emphasize that $T_c, \tilde{\mu}_c$ are not being used as fitting parameters. This  has eliminated a significant source of errors and allowed applying the FSS analysis \cite{FSS} to determine some critical exponents in a controlled manner within 1\% of the combined error.   However, as explained in Ref.\cite{Max}, the developed approach does not allow obtaining all the critical exponents because of the mixing effect  \cite{Mermin,Pokrovskii,Patashin,Bruce,Wilding}. Here we have developed  the NF method further. As a result, the mixing coefficient and the exponents $\mu, \nu$ (characterizing the divergence of the correlation length in the strong and weak field regimes, respectively) for 2D LG critical point are obtained within an error $\sim$ 1-2\%.  The hyperscaling relation is also confirmed. Thus, all other indices can be restored through the scaling relations (see in \cite{Landau}).   
These exponents are consistent with the Onsager solution for 2D Ising model. 

 It is also important to note that the position of the coexistence line and the mixing coefficient are found within the margin of error of only 0.1\%. This paves the way toward a controlled accurate determination of the non-analytical corrections to the diameter \cite{Mermin,Pokrovskii,Patashin,Bruce,Wilding,Fisher2000,Fisher2001,Fisher2003} without actually explicitly observing the LG coexistence. 

\section{Numerical Flowgram method and the mixing effect} 
According to the linear mixing theory\cite{Mermin,Pokrovskii,Patashin} it is not known along which path toward the critical point in the space of the primary scaling parameters  $(\tau,h)$ the system approaches the critical point.
The assumption is that there is a linear relation between $\delta T\equiv T-T_c,\, \delta \tilde{\mu}\equiv\tilde{\mu} - \tilde{\mu}_c$ and $\tau, h$. It is convenient to represent this relation for $\epsilon =1/T$ and $  \tilde{\mu}$ as
\begin{equation}\label{MX2}
\begin{aligned}
\tau \sim   \delta \epsilon + r   \delta \tilde{\mu} \\
h \sim  \delta \tilde{\mu}  + s \delta \epsilon
\end{aligned} 
\end{equation}
with some finite coefficients $s \neq r$. Here $\delta \epsilon \equiv \epsilon - 1/T_c = - \delta T/T_c^2$ and $\tau ,\, h$ stand for the parameters of the thermal and field operators, that is, as the deviations from $T_c$ and external field, respectively, in the $\varphi^4$ theory of scalar real field $\varphi$.

\begin{figure}[!htb]
	\includegraphics[width=1.0 \columnwidth]{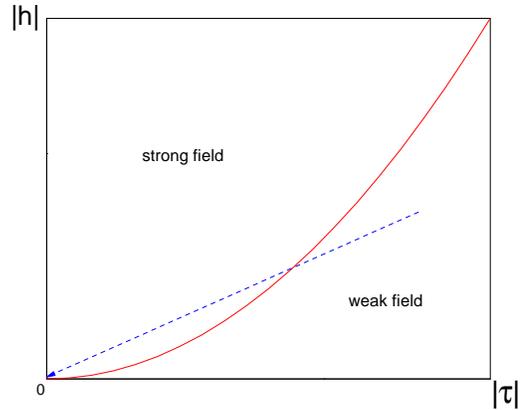}
	\vskip-8mm
	\caption{(Color online) A generic path (dashed line) toward the critical point ($h=0,\,\tau=0$) in presence of the mixing effect when $\mu <\nu$. The solid line, $h^* \sim |\tau|^{\nu/\mu}$  with  $\nu/\mu >1$, separates the regions of strong and weak field. }
	\label{fig:generic}
\end{figure}

The coexistence line corresponds to the $h=0$ condition. 
As discussed in Ref.\cite{Max}, a generic approach to the critical point is guaranteed to be dominated by the strong field behavior (see in \cite{Landau}) as long as the $\mu$ exponent, controlling the divergence of the correlation length $\xi \sim |h|^{-\mu} \to \infty $ as $h \to 0$ at $\tau=0$, is smaller than the $\nu$ exponent  
which determines $\xi \sim |\tau|^{-\nu} \to \infty$ as $\tau\to 0$ at $h=0$. To illustrate this aspect which is important to our present work, the sketch, Fig.~\ref{fig:generic}, features the (curved solid) line $|h|=h^* \sim |\tau|^{\nu/\mu}$ separating the regions of strong and weak field and a generic straight path (dashed straight line) toward the critical point $h=0, \tau=0$. As long as the slope of a path  is finite, it will eventually enter the strong field region close enough to the critical point. This implies that, generically, the critical behavior is controlled by the strong field exponents (see in Ref.~\cite{Landau}).  Conversely, there is only one path (with zero slope in Fig.~\ref{fig:generic}) toward the critical point which exists in the weak field region. Finding such a path and determining the critical behavior along it is the main purpose of this work.

 At this point it is worth briefly outlining how a critical exponent, say $\mu$, of the correlation length $\xi$ can be found by measuring some Binder cumulant \cite{Binder} $U_B$
and its derivative $dU_B/dt$ with respect to a control parameter $t$. The key point is to tune the system to the regime where $U_B$ belongs to the domain corresponding to critical behavior with $\xi > L$ for a given system size $L$. This domain is characterized by $U_1 <U_B<U_2$, where $U_{1,2}$ are values away from the criticality in the corresponding phases.
Plotting $dU_B/dt$  vs $U_B$ in this domain will give a family of self-similar curves characterized by the range $\sim L^{1/\mu}$ for a sequence of increasing $L$ values. Then, plotting the corresponding rescaling factor $\lambda(L)$ versus $L$ provides the $\mu$ exponent. This is the main approach of the numerical flowgram method \cite{Annals} applied to the LG critical point in Ref.\cite{Max}. It guarantees that the critical point is eventually reached in the limit $L\to \infty$.

The  standard approach for finding the coexistence line is based on collecting the histogram of density and observing the bimodal distribution. As explained above, this approach relying on the extrapolation of liquid and gas densities is the biggest source of errors.  
One option to find the path along the line $h=0$ without any extrapolation has been suggested in Ref.\cite{Max}. It is based on the drastically different divergence of the heat capacity $C\sim L^{\alpha/\nu}$ along the coexistence line compared to a generic path where $C\sim L^{\gamma/\nu}$. [ $\alpha, \gamma $ are the critical indices of the heat capacity and compressibility \cite{Landau}, respectively]. Normally, $\alpha$ is smaller than $\gamma$. Thus, there should be a well pronounced minimum in $C$  with respect to an angle $\phi$ of the path in the plane $( \epsilon, \tilde{\mu}) $. This procedure allows obtaining the angle $\phi$ corresponding to $h=0$ in Fig.~\ref{fig:generic} and also the $\nu$ exponent. More details are given below.
	
\section{Square well fluid model in 2D}

The square well potential is one of the simplest models that exhibits solid, liquid and gas phases \cite{Rotenberg}. The system of classical particles is described by the grand canonical  partition function

\beq
Z = \sum_{N=1}^\infty \frac{1}{N!}  e^{\tilde{\mu} N}\int d\vec{r}_1 ....d\vec{r}_N e^{- V},
\label{Y}
\eeq
where $V= (1/2)\sum_{ij} v(\vec{r}_i -\vec{r}_j)$ is the potential energy of binary interaction (normalized by temperature) between $N$ particles located at $\vec{r}_i, \, i=1,2,...N$ within the square area $L^2$; $\tilde{\mu}$ is the chemical potential normalized by temperature. 
The interaction energy  $v(\vec{r})$ between two particles separated by a vector $\vec{r}$ is taken as the square well potential.
That is, $     v =   \infty$ if $|\vec{r}|<\sigma$;  $ v= -\epsilon_0 <0 $ if $ \sigma \leq |\vec{r}|  \leq \tilde{\lambda}\sigma $; and $ v=0$ if $ r > \tilde{\lambda}\sigma$.
Here $\sigma$ and $\tilde{\lambda}\sigma > \sigma$ are the hard and soft core diameter, respectively, and $\epsilon >0$ characterizes attraction within the soft core shell. Since temperature is absorbed into the definition of $\epsilon$, we will be calling $1/\epsilon$ as``temperature" $T$ and $\tilde{\mu}$ as ``chemical potential". It worth mentioning that the energy $E= - \epsilon_0 N_p$ of the system is simply given by the total number of pairs $N_p$ of particles which are within the soft attraction radius of each other. The Monte Carlo simulations of the model have been conducted for $\lambda=1.5$ and $\sigma = 1.0$ using the Metropolis Algorithm by inserting and removing particles. 

 Using the flowgram method described above and the specific property of the Binder cumulant for LG system \cite{Fisher2003}, the critical point for the above model has been previously found in Ref.\cite{Max} as $T_c=0.5540\pm 0.0005$ and $\tilde{\mu}_c=-3.700 \pm 0.005$. The error is controlled by the maximal simulated size $L=84$ used to find the separatrix in $U_B$ corresponding to the critical point (as explained in Ref.\cite{Max}). We use these values to perform simulations in the vicinity of the critical point in order to determine the coexistence line and to measure the critical exponents.

 Here, we consider the following Binder cumulant \cite{Binder}
\begin{equation}
U_{4} = \frac{\langle (N-\langle N \rangle )^2 \rangle^2}{\langle (N-\langle N \rangle )^4 \rangle},
\label{U4}
\end{equation}
and its derivatives $dU_4/d\tilde{\mu}$ and $dU_4/d\epsilon$, where $\langle ... \rangle$ stands for the average with respect to the ensemble (\ref{Y}). These derivatives can be expressed in terms of the cumulants $\langle N^m N_p^k\rangle$, with $m,k=1,2,3,4,5$. 

A specific property of $U_4$ in the free space LG system has been addressed in Refs.\cite{Fisher2003}. The maximum of $U_4$ (where $dU_4/d\epsilon =dU_4/d\tilde{\mu}=0$) tends to $U_4=1$ along the coexistence line and to $U_4=1/3$ in a single phase, and these values are essentially independent of  $T,\tilde{\mu}$ in the respective domains. At the critical point, $U_4$ has a maximum reaching the scale independent value $U_4=U_c = 0.855  \pm 0.005$ (cf. $U_c=0.8562157(5) $ for square Ising model \cite{Blote}) -- thus forming a separatrix in the family of curves $U_4$ vs $L$ \cite{Max}. 
Consequently, the critical region (where $\xi >L$) can be identified by any value of $U_4$ different from $U_4=1$ and $U_4=1/3$. It is important to note that, as $L$ increases, the critical domain of  $T, \, \tilde{\mu}$ shrinks as $\sim L^{-1/\nu} \to 0$ or $\sim L^{-1/\mu}\to 0$ depending on how the critical  point is approached. In other words, if $T, \tilde{\mu}$ are tuned to keep $U_4$ within its critical range for all simulated sizes, it is guaranteed that the system is critical for a given size $L$. This means that any quantity demonstrating scaling behavior will scale as a power of $L$ determined by its scaling dimension -- if plotted vs $U_4$. 

\subsection{Critical behavior in the strong field regime}
Within the field theory of $Z_2$ criticality the FSS behavior of heat capacity $C$ is insensitive to the path toward the critical point $\tau=0,\, h=0$. In the weak and strong field regions $C\sim L^{\alpha/\nu}$ and $C\sim L^{\varepsilon/\mu}$, respectively. However, the scaling relations guarantee that $ \alpha/\nu= \varepsilon/\mu$ (see e.g. in \cite{Landau}).   In 2D Ising model $\alpha=\varepsilon=0$ (which implies log-divergence $C \sim \log L$). 
The situation is different in the case of the LG critical point due to the mixing effect. 
Along a path toward the critical point belonging to the strong field region energy $E$ and particle number $N$ fluctuations are linearly mixed. Thus, the divergence of heat capacity $C=-d^2 \ln Z/d \epsilon^2$ (defined up to a constant factor) is controlled by the much stronger divergence of compressibility. Within the FSS this gives $C \sim  L^{(1-1/\delta)/\mu}=L^{\gamma/\nu}$, where $\delta$ is the critical index determining  order parameter behavior in the strong field region and $\gamma$ is the index of compressibility (magnetic susceptibility) in the weak field regime (see in Ref.\cite{Landau}).    In 2D this gives $C\sim L^{7/4}$.
\begin{figure}[]
	\includegraphics[width=1.0 \columnwidth]{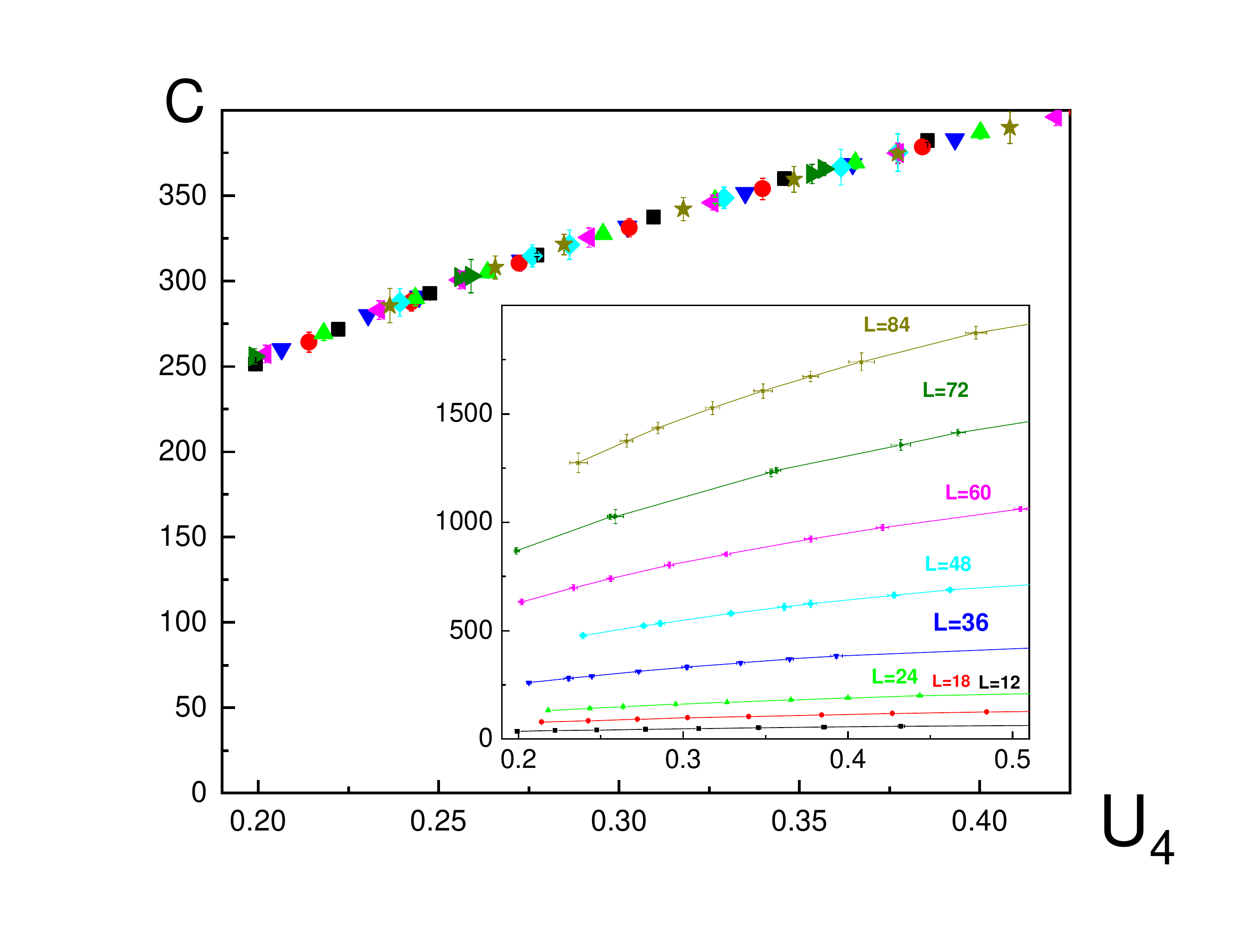} 
		\vskip-8mm
	\caption{(Color online) The master curve of the specific heat $C$ vs $U_4$ for various system sizes $L$ obtained by rescaling  particular curves along the "vertical" direction until each data set overlaps with the curve corresponding to $L=36$. Inset: The non-rescaled data of $C$ vs $U_4$ for sizes shown close to each curve. }
	\label{fig:Cv-mu}
\end{figure}
\begin{figure}[]
	\includegraphics[width=1.0 \columnwidth]{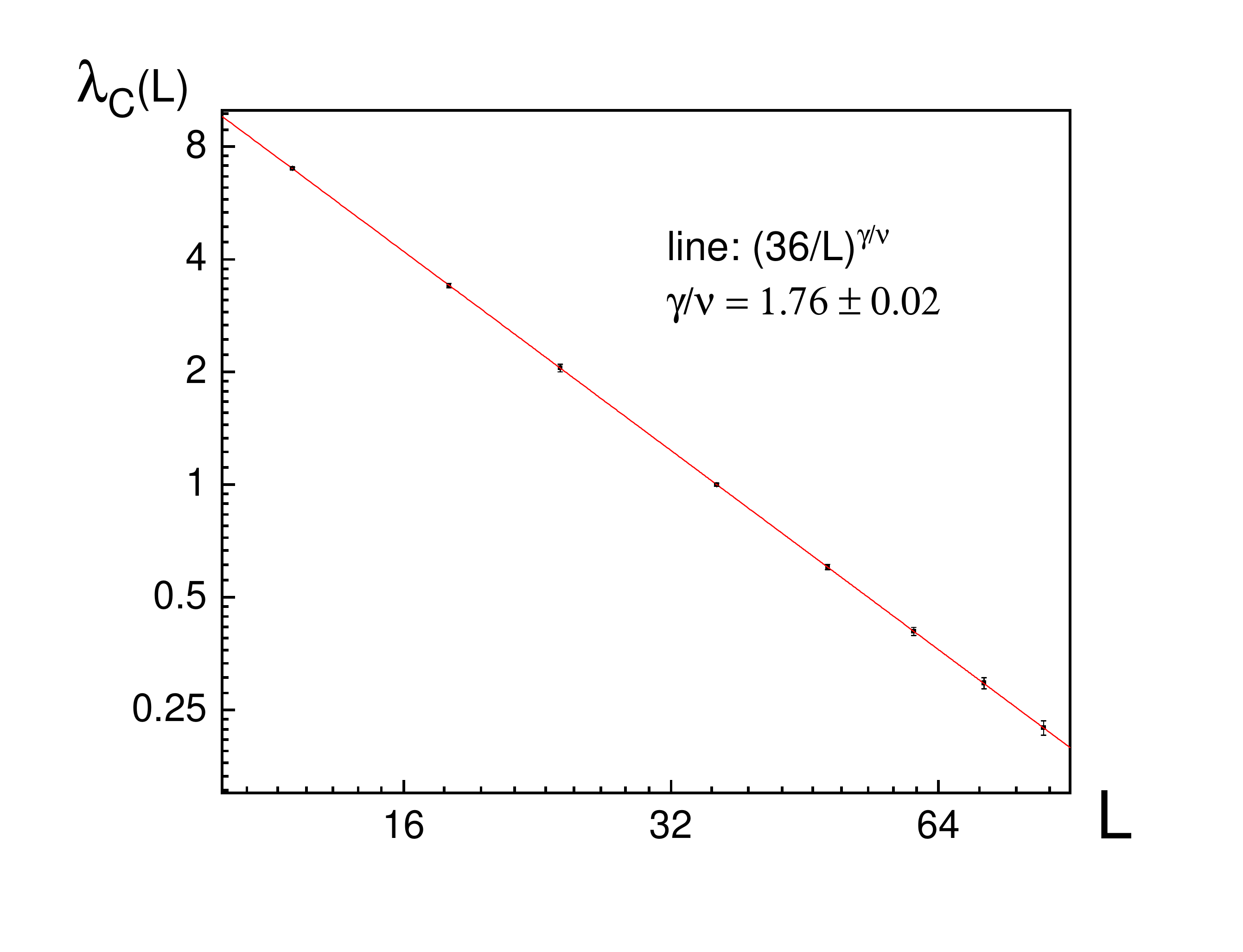} 
		\vskip-8mm
	\caption{(Color online) Rescaling factor $\lambda(L)$ versus $L$ for the data shown in Fig.~\ref{fig:Cv-mu}. The slope of the fit line (solid straight line) gives the exponent $(1-1/\delta)/\mu=\gamma/\nu=1.76 \pm 0.02$ which is consistent with the Onsager value $\gamma/\nu =7/4=1.75$.}
	\label{fig:log-log-cv}
\end{figure}

\begin{figure}
	\centering
	\includegraphics[width=1.0 \columnwidth]{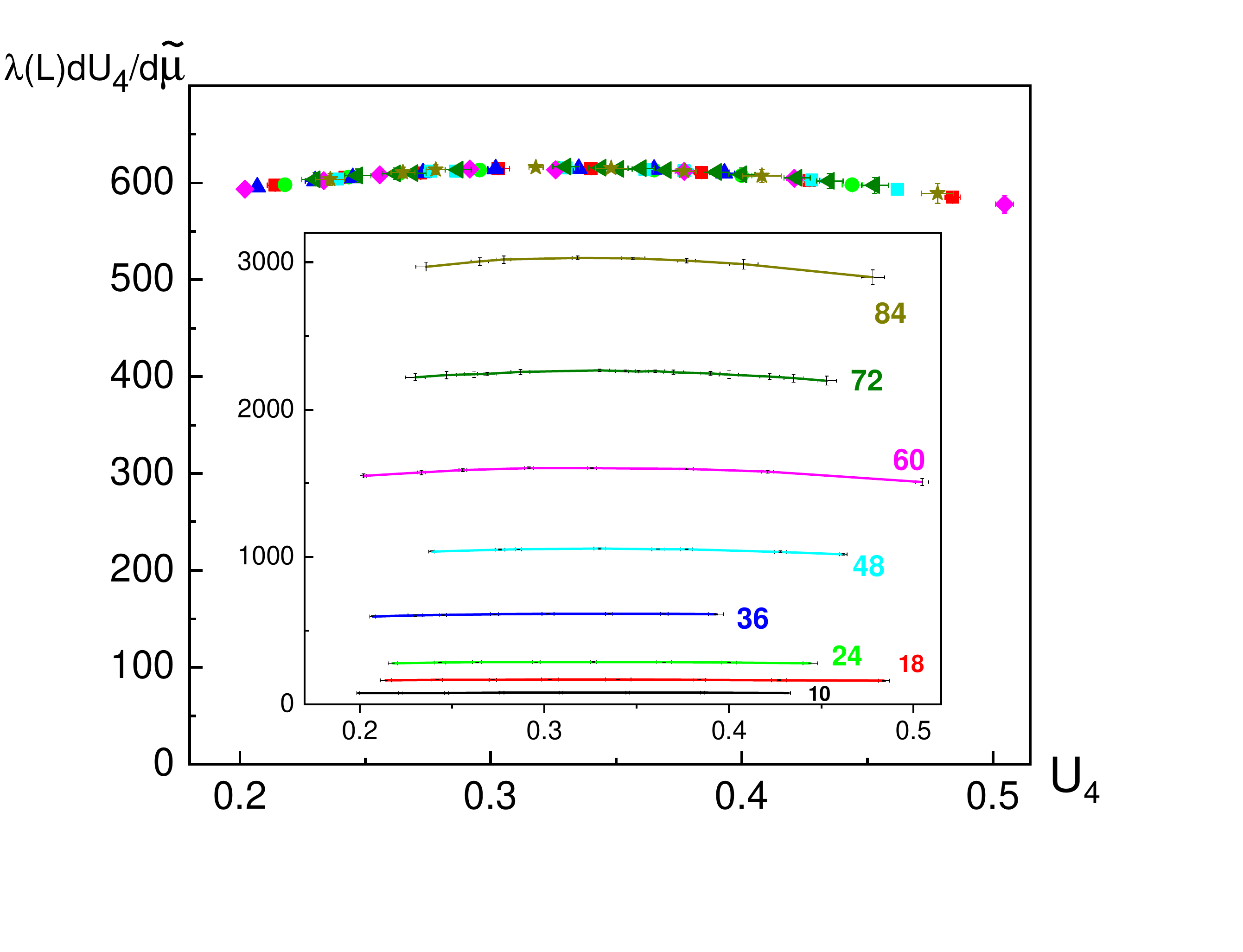}
		\vskip-8mm
	\caption{(Color online) The master curve of  $\frac{dU_4}{d\tilde{\mu}}$ vs $U_4$, Ref.\cite{Max}, for various system sizes $L$ obtained by the "vertical" rescaling
to the curve corresponding to $L=30$. Inset: the original curves for sizes shown close to each data set.}
	\label{pi2}
\end{figure}

\begin{figure}
	\centering
	\includegraphics[width=1.0 \columnwidth]{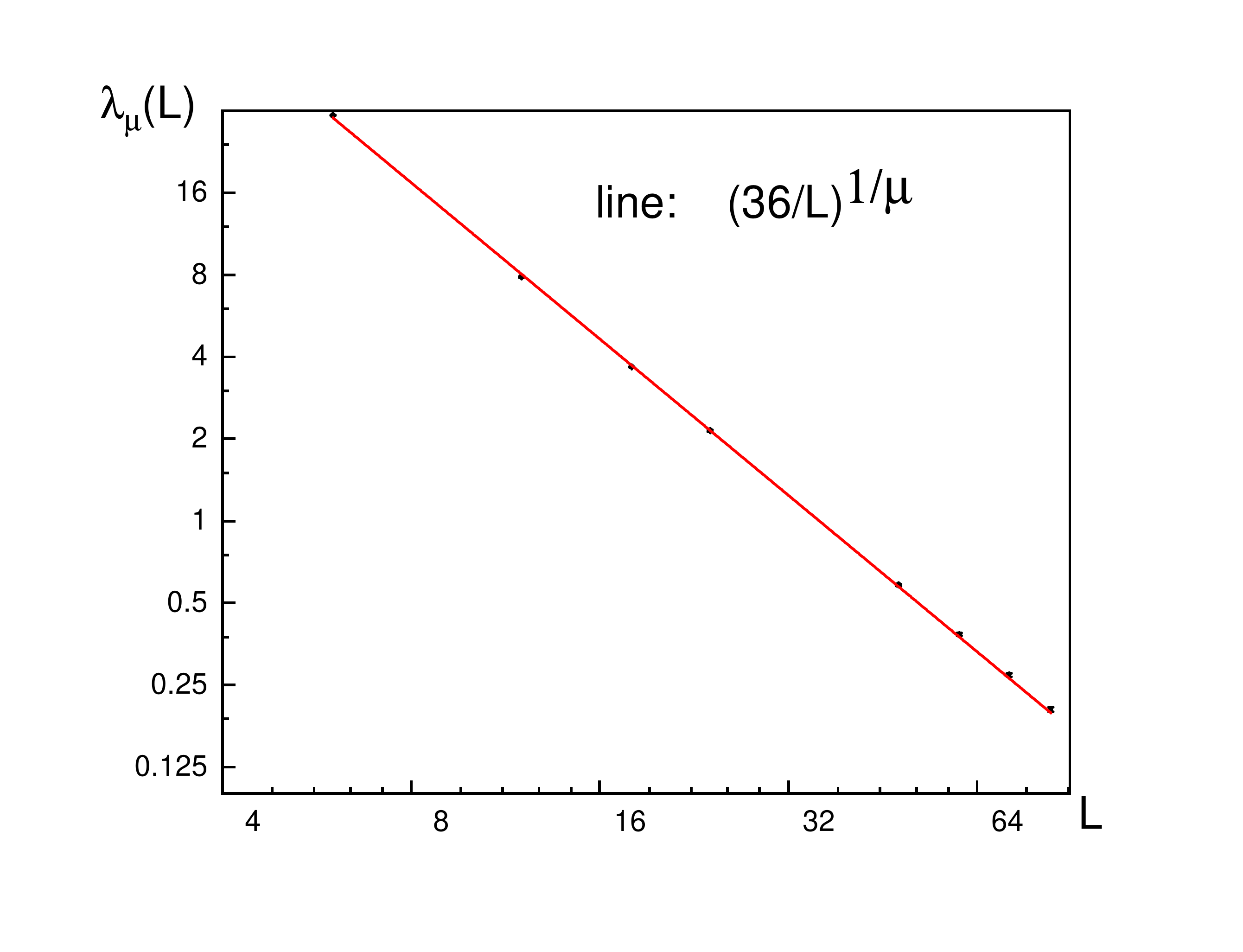}
		\vskip-8mm
	\caption{(Color online) The log-log plot of the rescaling parameter from the data shown in Fig.~\ref{pi2}. The slope of the fit line (solid straight  line) gives the $\mu\approx 0.535\pm  0.005$ exponent consistent with the Onsager value 8/15. }
	\label{lambda_mu}
\end{figure}

Fig.~\ref{fig:Cv-mu} displays the results of measuring the heat capacity $C$ along the path $(\delta T=0,\, \delta \tilde{\mu})$ obtained within the  flowgram method. The value of $\delta\tilde{\mu}$ was adjusted in such a way that $U_4$ falls into its critical range for each size $L$. For large enough $L$ the function $C$ vs $U_4$ is universal up to a scaling factor. 
The ``vertical" rescaling of $C$ for various $L$ by a factor $\lambda(L)\sim L^{-(1-1/\delta)/\mu}$ allows obtaining the exponent which turns out to be consistent with the Onsager value $7/4$ as indicated in  Fig.~\ref{fig:log-log-cv}.  [The error $\sim 2\%$ includes the subdominant scaling contribution].   
Similarly, derivative of $U_4$ with respect to either $\tilde{\mu}$ or $\epsilon$ at the same path demonstrates the strong field scaling $dU_4/d\tilde{\mu} \sim dU_4/d\epsilon \sim 1/h \sim L^{1/\mu}$ in the critical domain of $U_4$ \cite{Max}.
 This behavior is demonstrated in Figs.~\ref{pi2},\ref{lambda_mu} where the critical exponent $\mu$ was found in Ref.\cite{Max} to be consistent with the Onsager value $8/15$ within 1\% of the total error. 

\subsection{Critical behavior in the weak field regime}\label{weak}
\begin{figure}
	\includegraphics[width=1.0 \columnwidth]{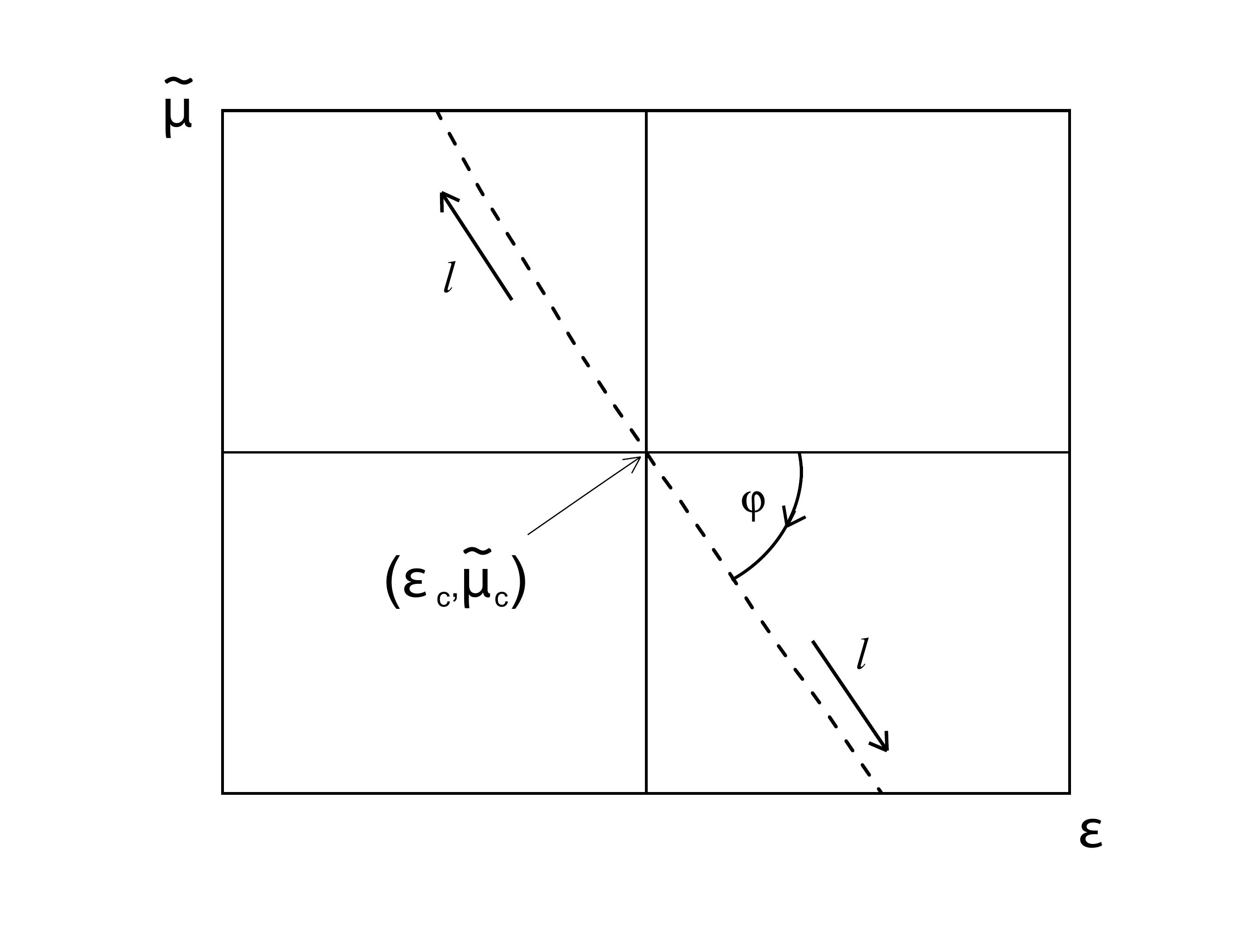}
	\caption{(Color online) Polar representation of the system parameters with $\epsilon_c = 0.5540$ and $\tilde{\mu}_c = -3.701$ as found in Ref.\cite{Max}. The dashed line in the quadrant $\delta \epsilon >0, \, \delta \tilde{\mu} <0$ marks the coexistence line ending at the critical point $\delta \epsilon =0,\, \delta \tilde{\mu}=0$.  } 
	\label{fig:polar}
\end{figure}

As mentioned  above, there should be a minimum of $C$ as a function of the path slope with respect to the axis $\epsilon$ (or $\tilde{\mu}$). This minimum corresponds to the coexistence line $h=0$. 
It is convenient to represent the deviations of the system parameters $\delta \tilde{\mu}, \delta \epsilon$ from the critical point in polar coordinates $(l,\phi):$  
\begin{equation}\label{polar}
\delta \tilde{\mu}=-l\sin\phi,\quad \delta \epsilon= l\cos\phi
\end{equation}
as depicted in Fig.~\ref{fig:polar}.
The coexistence line is characterized by some angle $\phi=\phi_V$ in the quadrant $\delta \epsilon >0,\, \delta \tilde{\mu}<0$ in Fig.~\ref{fig:polar}.
Comparison  with Eq.(\ref{MX2}) gives
\begin{equation}
s=\tan \phi_V.
\label{sr}
\end{equation}

We introduce the derivative $C =- d^2\ln Z/dl^2/L^2$ at fixed angle $\phi$. Keeping in mind that $Z\sim \sum \exp( - E\epsilon + \tilde{\mu} N)$ and also Eq.(\ref{polar}), we find  
\begin{multline}\label{C}
C= \frac{1}{L^2}[\cos^2\phi\langle\delta E  ^2  \rangle +  \sin^2\phi\langle  \delta N  ^2  \rangle +  \sin(2\phi) \langle \delta N \delta E\rangle] ,
\end{multline}
where $\delta E,\, \delta N$ are fluctuations of $E$ and $N$, respectively. 
This derivative can be viewed as the specific heat along the path determined by a fixed angle $\phi$ toward the critical point.
It is useful to represent $C$ in terms of Eqs.(\ref{MX2},\ref{polar}) used in the free energy representation $F(\tau, h)$ in terms of the primary
parameters as $F =F(A l, B l)$, where
$A\equiv \cos \phi -r \sin \phi,\,\, B\equiv  - \sin \phi + s \cos \phi$. Then, the differentiation gives
\begin{multline}\label{Ctau}
C=\frac{1}{L^2}\left[  A^2 \frac{d^2F}{d\tau^2} +2AB\frac{d^2F}{d\tau dh} + B^2 \frac{d^2F}{dh^2}\right].
\end{multline}
  For finite $B$ the dominant term in Eq.(\ref{Ctau}) is the last one showing the divergence $\sim \tau^{-\gamma}$ of the magnetic susceptibility \cite{Landau}  . This term as well as 
the second one cancel out for $B=0$ which results in the much weaker singularity $C \sim d^2 F/d\tau^2 \sim \tau^{-\alpha}$. As can be seen, the condition $B=0$ coincides with Eq.(\ref{sr}). 
  As explained in Ref.\cite{Max} and also above, this implies that $C$ vs $\phi$ reaches minimum at $\phi=\phi_V$, and this is the feature which is used here to determine the coexistence line. It is important that the minimum becomes progressively more pronounced as $L$ increases, provided $U_4$ stays in its critical domain.

\begin{figure}
	\includegraphics[width=1.0 \columnwidth]{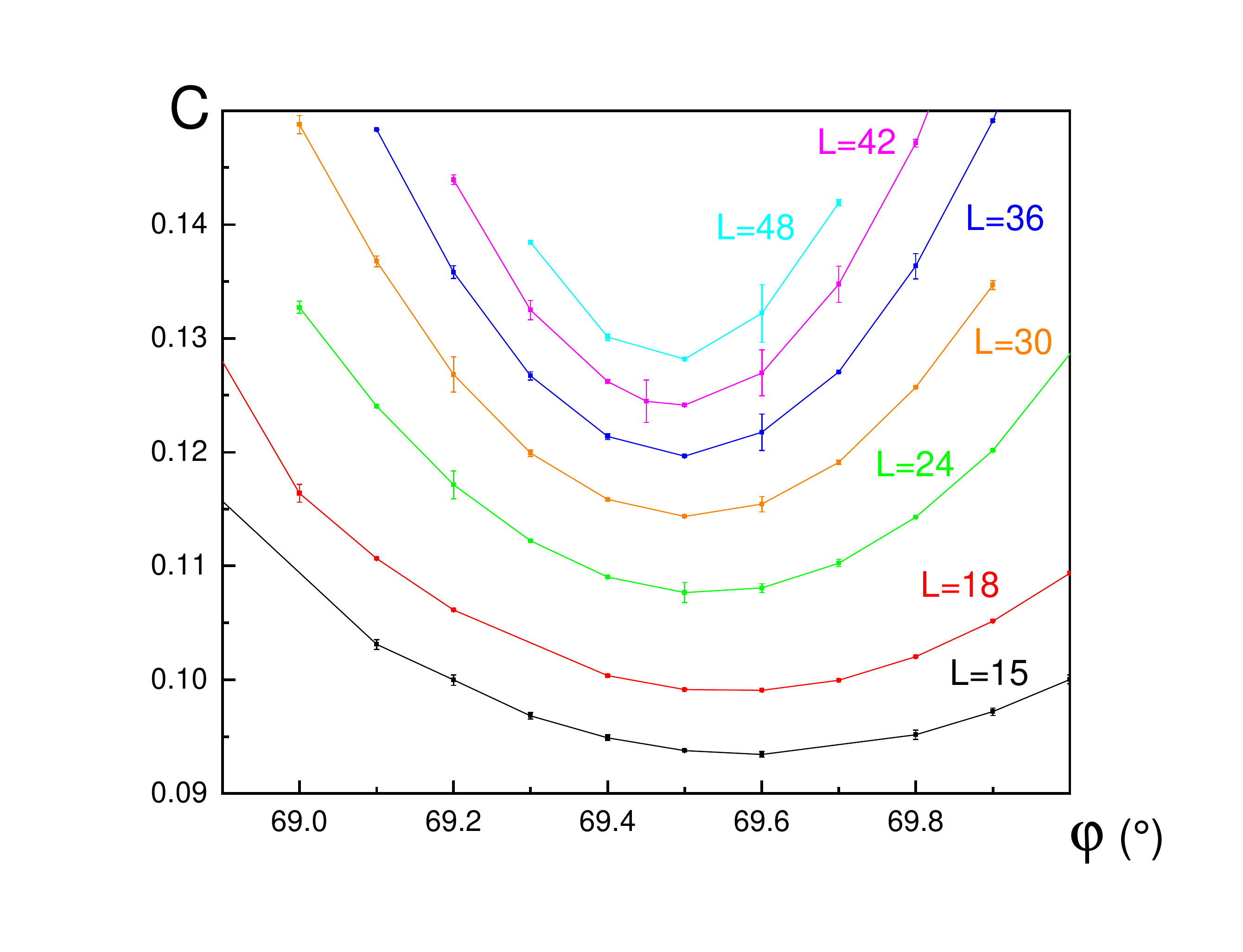}
		\vskip-8mm
	\caption{(Color online) The specific heat along the path for $l=10^{-6}$ guaranteeing that $U_4 = U_c$ within 1-2\% of deviations in the critical region surrounding the minimum at $\phi=\phi_V$.}
	\label{fig:d2Fdl2}
\end{figure}
\begin{figure}
\centering
	\includegraphics[width=1.0 \columnwidth]{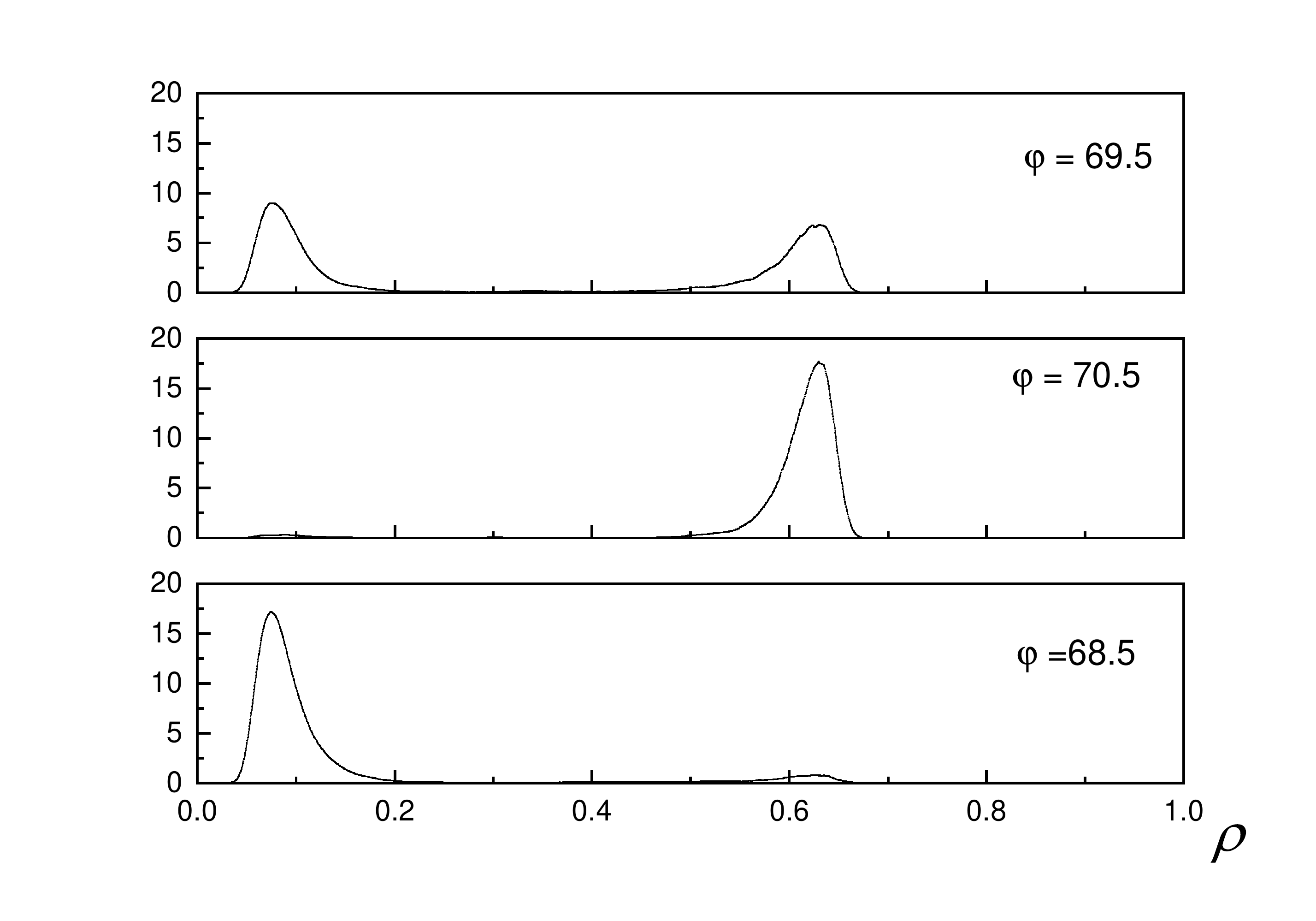}
		\vskip-8mm
	\caption{Histograms of density along the straight path toward the critical point, Fig.~\ref{fig:polar}, for three angles.
  }
	\label{fig:hist}
\end{figure}
\begin{figure}
	\includegraphics[width=1.0 \columnwidth]{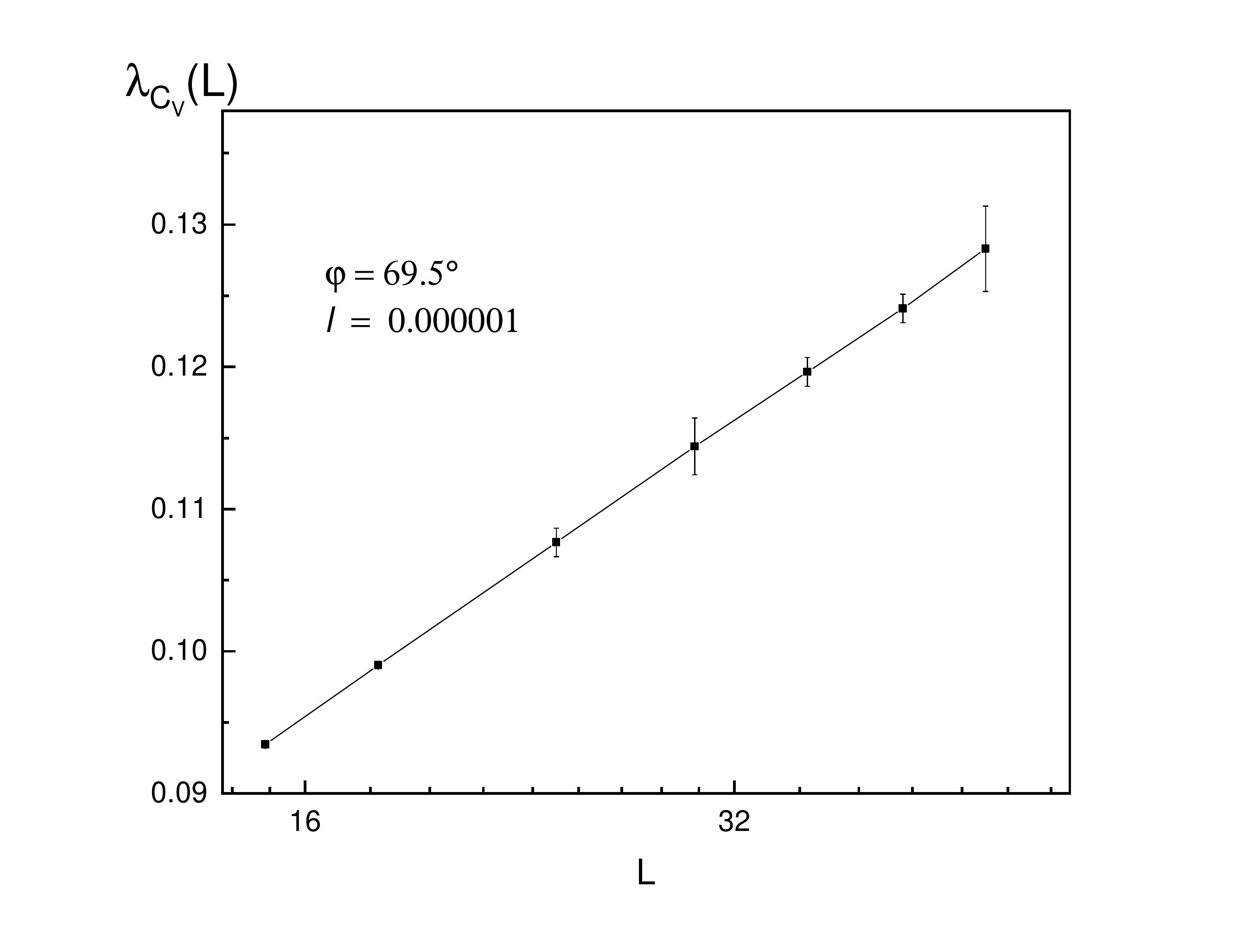}
		\vskip-8mm
	\caption{(Color online) The minima of $C_V$ shown in Fig. \ref{fig:d2Fdl2} vs $\log L$. }
	\label{fig:log-d2Fdl2}
\end{figure}

In order to find $\phi_V$ measurements of $C=C(\phi)$, Eq.(\ref{C}), were conducted for various $\phi$ around the minimum of $C$ vs $\phi$  for several system sizes.  Simulations were conducted at fixed $l$ for each size.
 Fig. \ref{fig:d2Fdl2} displays the results of such measurements giving the angle $\phi_V = 69.5\degree\pm 0.1\degree$ .
In order to demonstrate that the path in Fig.~\ref{fig:polar} at $\phi=\phi_V$ corresponds to the coexistence line, Fig.~\ref{fig:hist}
shows three histograms of the density collected at $\phi=\phi_V$ and at slightly different angles.   The top panel, where $\phi=\phi_V$, features the bimodal distribution corresponding to LG coexistence. The middle and the lower panels represent liquid and gas, respectively. As another crosscheck, the heat capacity $C$ at $\phi=\phi_V$ is plotted in Fig.~\ref{fig:log-d2Fdl2}. Its $L$ dependence is consistent with 
 $C_V \sim \log L$, that is, $\alpha =0$.  

\begin{figure}
	\centering
	\includegraphics[width=1.0 \columnwidth]{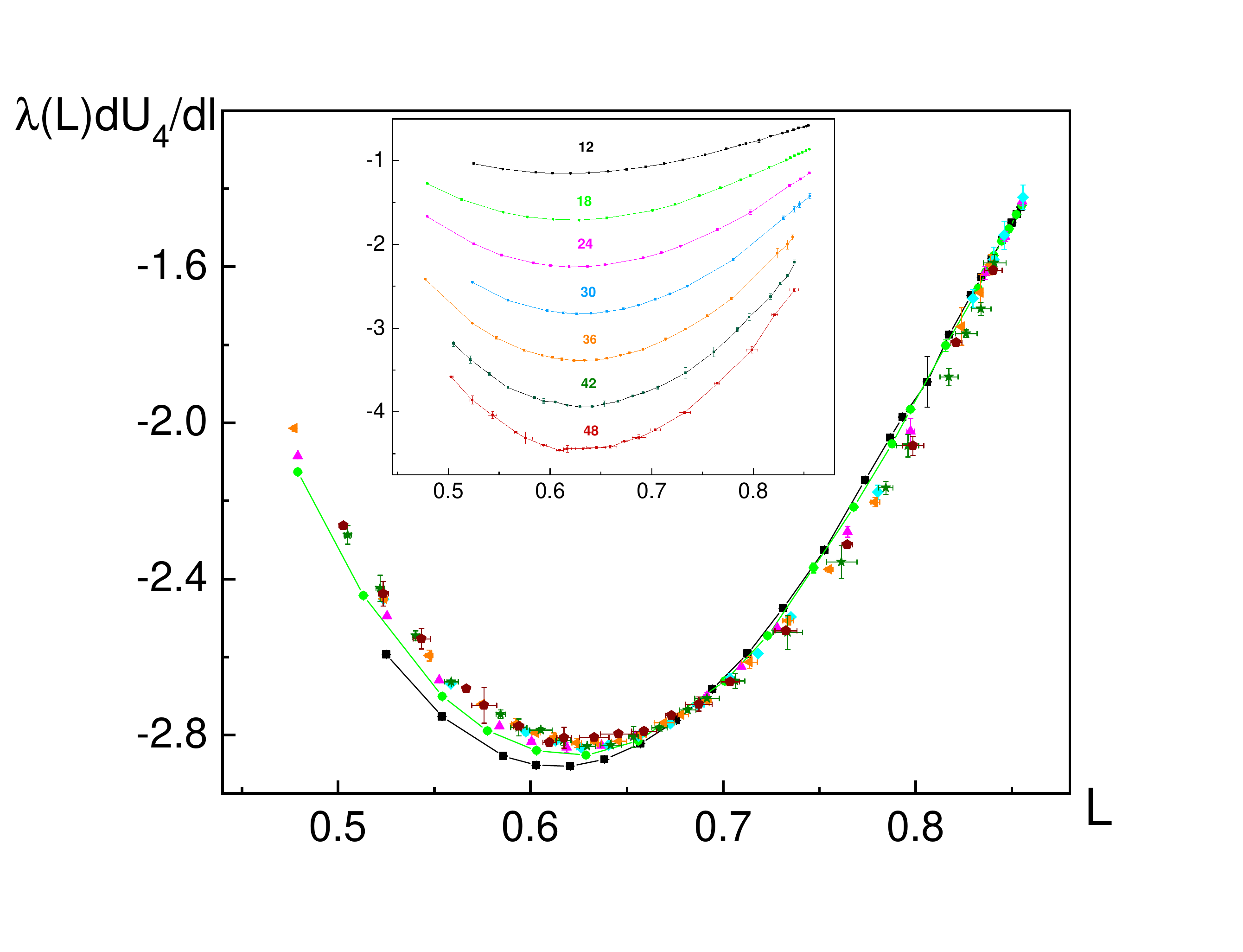}
		\vskip-8mm
	\caption{(Color online) The master curve of  $\frac{dU_4}{dl}$ vs $U_4$ at  $\phi=\phi_V=(69.5 + 180)\degree$  (that is, along the continuation of the coexistence line beyond the critical point in Fig.~\ref{fig:polar}) for various system sizes $L$ obtained by the "vertical" rescaling
to the curve corresponding to $L=30$. The data points for $L=12,18$ are connected by lines as an indication that these sizes fall out from the master curve. Inset: the original curves for sizes shown close to each data set.}
	\label{fig:dU4dl}
\end{figure}

\begin{figure}
	\includegraphics[width=1.0 \columnwidth]{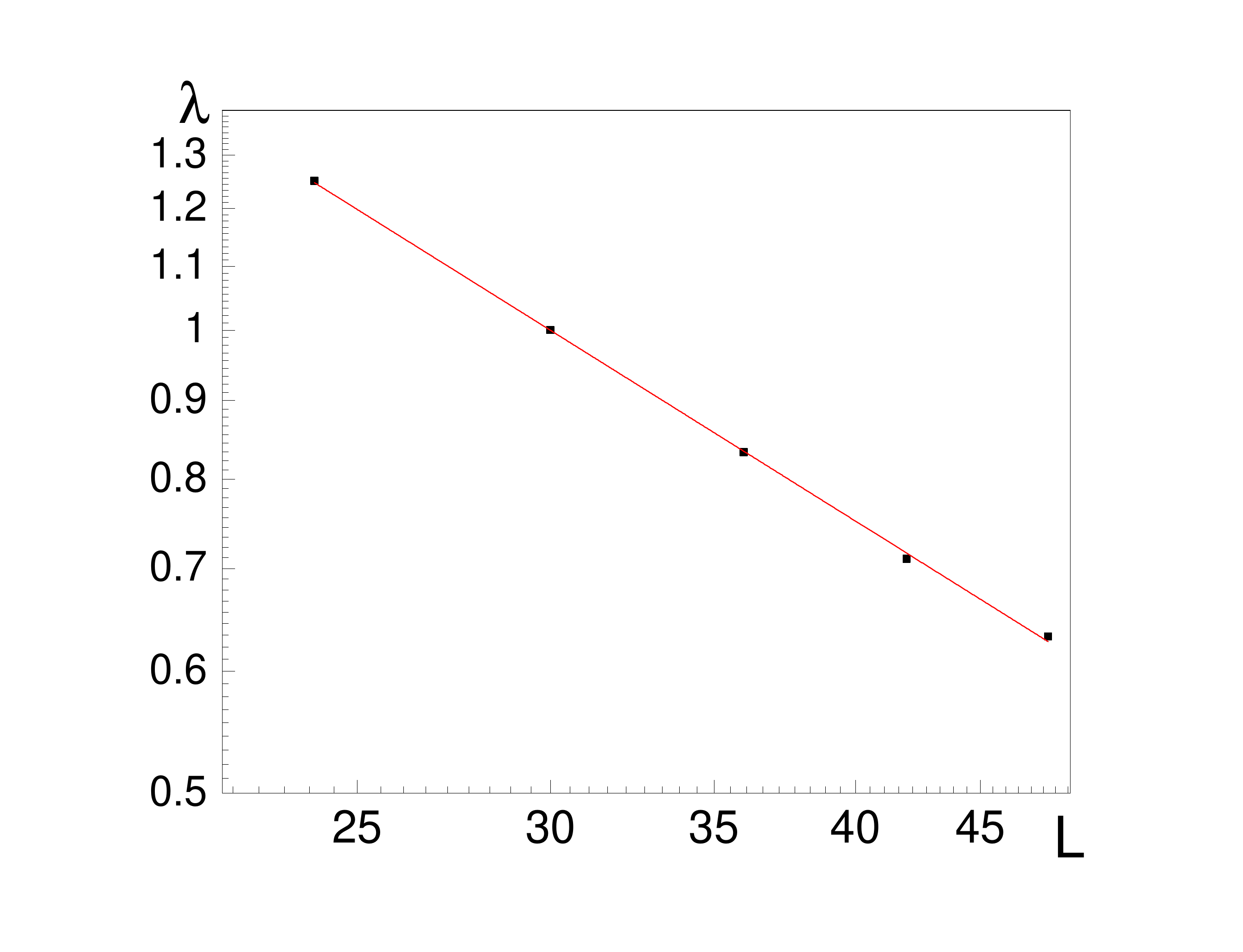}
		\vskip-8mm
	\caption{(Color online) Log-log plot of the rescaling parameter $\lambda (L)$ of $dU_4/dl $ from Fig. \ref{fig:dU4dl}, with the smallest sizes $L=12,18$ excluded. The slope of the straight fit line gives $\nu=1.01$ with the fit error of 1\% .}
	\label{fig:dU4dl-log-log}
\end{figure}

The quantity $l$ at $\phi=\phi_V$ plays the role of the thermal operator parameter $\tau$. Thus, its scaling dimension must be $1/\nu$.
In other words, the derivative $\frac{dU_4}{dl}$ , which can be represented as
\begin{equation}\label{Du4X}
\frac{dU_4}{dl} =\cos \phi_V [ \frac{dU_4}{d\epsilon} - \tan \phi_V \frac{dU_4}{d\tilde{\mu}}] 
\end{equation}
with the help of Eq.(\ref{polar}),
vs $U_4$ must scale as $\sim L^{1/\nu}$ in the critical domain. 
The family of such curves, that is, $dU_4/dl$ vs $U_4$ is shown in Fig.~\ref{fig:dU4dl}. The data points have been collected along the continuation of the coexistence line beyond the critical point -- that is, at $\phi = (69.5 +180)\degree $ -- in the quadrant $\delta \tilde{\mu}>0, \, \delta \epsilon <0$. [ This path corresponds to $\tau >0$ in the $\phi^4$ theory]. As can be seen,
the curves can be collapsed to a master curve by the ``vertical" rescaling for sizes above $L=18$.
The log-log plot of the rescaling parameter vs $L$ gives the critical exponent $\nu=1.01 \pm 0.02$ as shown in Fig.~\ref{fig:dU4dl-log-log}.  
The error consists of 1\% of the statistical error and 1\% of the fitting error for the data points excluding sizes $L=12,18$. [The minimal statistical error of less than 1\% characterizes  the data points close to the minimum of the master curve -- that is, in the domain $0.55< U_4<0.75$].  The smallest sizes fall out from the collapse due to the subdominant scaling contribution.   

\subsection{Covariance at the coexistence line}
\begin{figure}
	\includegraphics[width=1.0 \columnwidth]{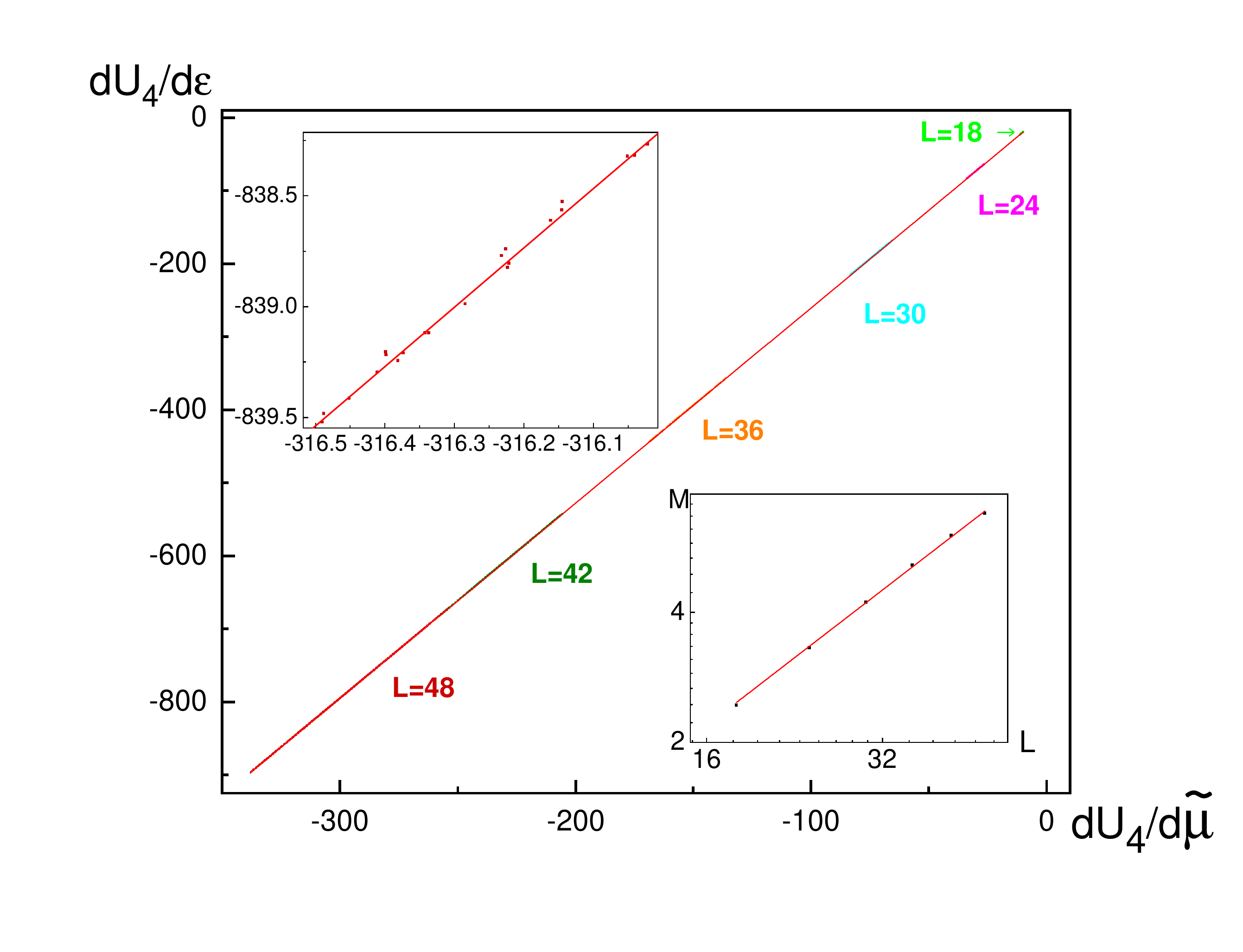}
		\vskip-8mm
	\caption{(Color online) Covariance plot of $\frac{dU_4(S)}{d\epsilon}$ vs  $\frac{dU_4(S)}{d\tilde{\mu}}$ for $L=18, ..., 42$ collected at the coexistence line at $l=10^{-6}, \phi=\phi_V=69.5\degree$. The data for different sizes are shifted by the intercept difference with the $L=48$ data. The points are shown 
for every $\Delta S \sim 10^9-10^{10}$ steps for MC runs lasting a total of $10^{12}\sim 10^{13}$ MC steps. The size of the data points is larger than the
deviations from the fit (solid) line. The upper inset: the part of the $L=48$ covariance data is shown to indicate typical deviations  from the fit line which has the slope $K=2.674 \pm 0.002$.
Lower inset: The log-log plot of the intercept $M$ vs $L$. Its slope $1/\nu$ yields $\nu =0.96 \pm 0.05$.}
\label{covariance}
	\vskip-5mm
\end{figure}

In lattice gas models with underlying $Z_2$ symmetry fluctuations of energy $\delta E$ and number of particles $\delta N$ are statistically independent, $\langle \delta E \delta N\rangle =0$ \cite{Patashin}. In free space systems, where there is no such $Z_2$ symmetry, these quantities are strongly correlated, that is, $\langle \delta E \delta N\rangle \neq 0$ \cite{Patashin} -- as determined by the mixing effect.    
Such correlations are revealed in the covariance plot of $dU_4/d\epsilon$ vs $dU_4/d\tilde{\mu} $, Fig.~\ref{covariance}.

The idea of covariance \cite{Sandvik} between two quantities $X(S)$ and $Y(S)$ is based on collecting incomplete statistical averages -- as  functions of a number $S$ of Monte Carlo (MC) steps
and then plotting $Y$ vs $X$ with $S$ considered as a parameter. The full statistical averages correspond to the limit $S \to \infty$. 
Strong correlations between $X$ and $Y$ can be revealed in a well defined curve $X=X(S),\, Y=Y(S)$ determined parametrically in the coordinate plane $(X,Y)$.

\begin{figure}
	\includegraphics[width=1.0 \columnwidth]{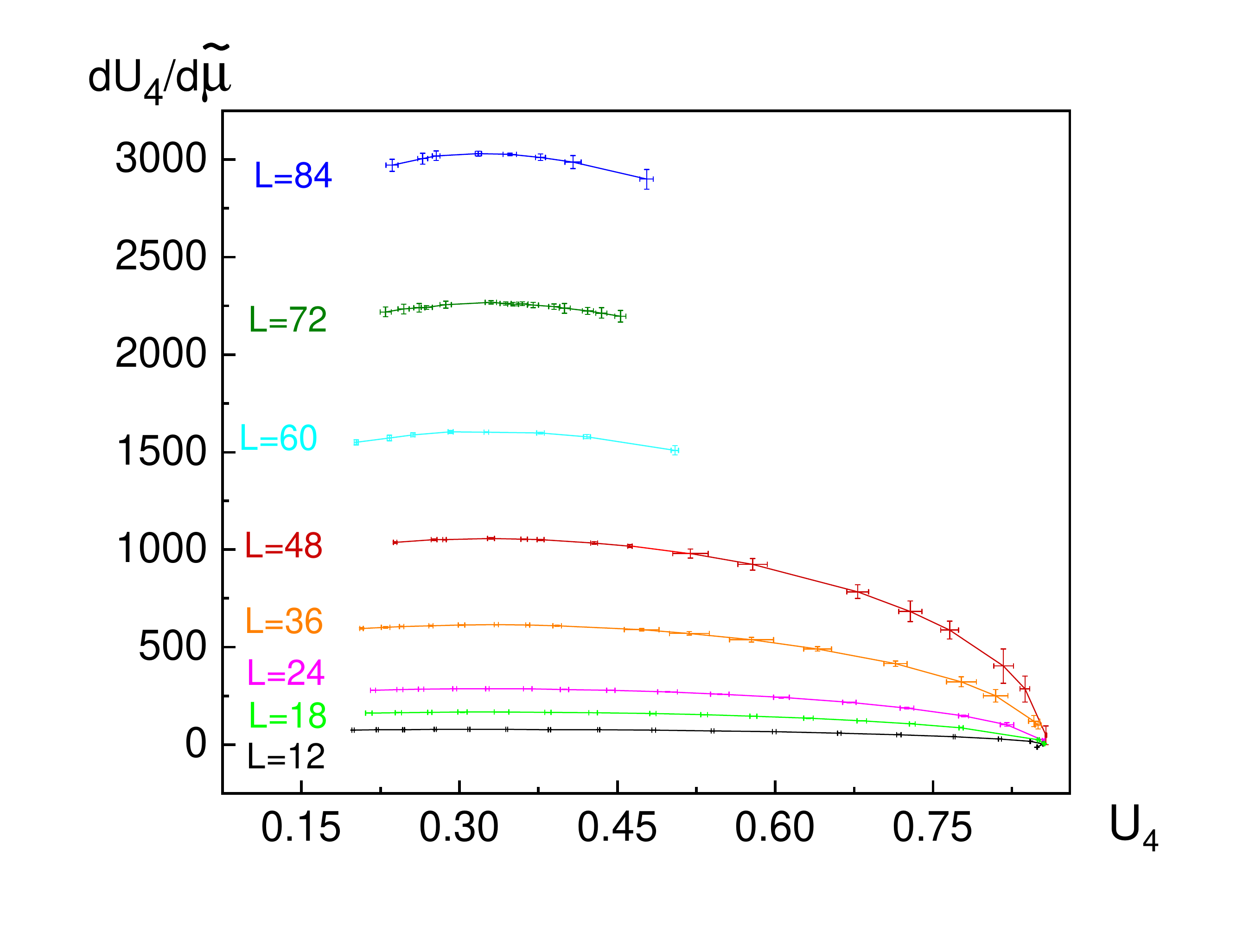}
		\vskip-8mm
	\caption{(Color online) The raw data for $dU_4/d\tilde{\mu}$ vs $U_4$ over the whole critical domain of $U_4$. It includes 
the data shown in Fig.~\ref{pi2}. }
\label{full}
	\vskip-5mm
\end{figure}
In our case each point in Fig.~\ref{covariance} corresponds to $10^{11}-10^{13}$ MC updates for the {\it same} system parameters $l, \phi$ and for different sizes $L$.
As can be seen, there are large fluctuations of $Y=dU_4/d\epsilon$ and $X=dU_4/d\tilde{\mu} $ and, simultaneously, these are characterized by a strong correlation between both quantities which form a well defined straight line  $Y = K X+ M$
with some  coefficients $K,M$. The extended domain and range of $X,Y$ are a direct reflection of the diverging critical fluctuations of the number of particles.

According to the FSS \cite{FSS}, the cumulant $U_4$ close to the critical point, where $\xi >L$, can be represented as an analytical  function of two variables $\tau,h$ in the form
$f(x,y)$ where $x =\tau L^{1/\nu},\, y=h L^{1/\mu}$. This function is universal and is characterized by $f(0,0)=U_c\approx 0.856$. It is symmetric with respect to $h \to -h$, so that exactly at the coexistence line $f'_y=\partial f/\partial y=0$.

 We introduce the incomplete statistical mean $U_4(\tau, h, S)$ of the Binder cumulant $U_4$ as well as its derivatives $Y(S)\equiv dU_4(\tau, h, S)/d\epsilon , \, X(S)\equiv dU_4(\tau, h, S)/d\tilde{\mu} $. These can be expressed in terms of the derivatives with respect to the primary parameters $\tau,\, h$ as 
\beq
 Y=\frac{dU_4}{d\tau} + s \frac{dU_4}{dh} , 
\label{covY}\\
 X=r  \frac{dU_4}{d\tau} +  \frac{dU_4}{dh} .
\label{covX}
\eeq
where Eq.(\ref{MX2}) has been utilized.  
Exactly at the coexistence line, $h=0$, the term $dU_4/dh= f'_y(x,y=0,S)L^{1/\mu}$  fluctuates strongly around $=0$ as a consequence of strong fluctuations characterizing order parameter in $Z_2$ field theory.  In the weak field region, where $h$ is finite but is not dominant with respect to $\tau$, the fluctuations remain strong and shifted with respect to the full statistical value (in the limit $S\to \infty$).   At the same time, the derivative $ \frac{dU_4}{d\tau}$ does not fluctuate strongly in the weak field region so that it can be replaced by its full statistical mean.
Expressing $Y$ in terms of $X$ we find that
 the covariance line in Fig.~\ref{covariance} has the slope $K=s$  and the intercept $M=(1-sr) L^{1/\nu} \frac{\partial f(x,y=0)}{\partial x} \sim L^{1/\nu}$, where we considered the weak field limit $h=0$. 
Thus, 
using Eqs.(\ref{covY},\ref{covX}) in Eq.(\ref{Du4X}) the derivative along the coexistence $dU_4/dl \sim dU_4/d\tau$ line can be expressed as  
\begin{equation}\label{Du4l}
\frac{dU_4}{dl} = \cos \phi  M \sim L^{1/\nu}.
\end{equation}
Below the critical point $\phi=\phi_V$ and above it $\phi=\phi_V + 180\degree$ (on the linear extension of the coexistence line). 
We should mention that $dU_4/dl$ in Fig.~\ref{fig:dU4dl} is shown above the critical point (that is, $\tau>0$ or $\phi = \phi_V +180\degree$) while the covariance plot, Fig.~\ref{covariance}, was obtained below the critical point ($\tau<0$ or $\phi=\phi_V$).

The measured value of the slope $K=s= 2.674 \pm 0.002 $ is consistent with the value of $s$, Eq.(\ref{sr}), found by the method discussed in the previous section.
It is worth mentioning that this
slope remains unchanged up to the value $l=0.1$ where the deviations become visible in the third digit ($K=2.69$). It is also important to note that the dominant error is due to the systematic variations $\sim 0.1\%$ between different $L$, with the statistical errors being at least two orders of magnitude lower.  
The scaling of the intercept $M$ has been determined to be consistent with $M\sim L^{1/\nu}$ where $\nu=0.96 \pm 0.05$.

At this point we notice that the margin of error in $\nu$ determined from the intercept $M$ is about 2-3 times bigger than the error obtained from the method discussed in Sec.~\ref{weak}.  The reason for such a difference lies in very different values of $M$ and the derivatives $dU_4/d\epsilon, \, dU_4/d\tilde{\mu} $ contributing to the slope $K$ in Fig.~\ref{covariance}. The intercept values $M$ are a factor of 100-200 smaller (see the lower inset in Fig.~\ref{covariance}) than the values of the derivatives. This has enhanced the relative error of $M$ and, consequently, of $\nu$.  
   
Finally, we discuss the impact of the covariance on statistical errors. Diverging critical fluctuations of $N$ affect the rate of convergence of simulations. Close to the critical point, where $U_c \approx 0.856$ and the derivatives of $dU_4/d\epsilon, dU_4/d\tilde{\mu}$ approach zero, the relative error becomes large and determining the scaling behavior requires much longer simulation times. This feature can be seen from Fig.~\ref{full} which includes the data shown in Fig.~\ref{pi2} for $U_4$ up to $U_4=0.5$ and also for $ 0.5< U_4 <U_c\approx 0.856$. 
The error bars in the second domain are significantly larger than in the first one used to measure the $\mu$ exponent -- to such an extent that presenting the data for $L>48$ becomes useless. This is a direct consequence of the diverging fluctuations of $N$ and $E$. However, it is important to emphasize that these fluctuations cancel out from $dU_4/dl$ measured along the coexistence line -- thanks to the covariance effect. [If the derivatives $dU_4/d\tilde{\mu}, dU_4/d\epsilon$ were collected independently from each other and, then, used in Eq.(\ref{Du4X}), the error bars would be more than order of magnitude larger]. Thus, an accurate determination of the $\nu$ exponent becomes possible from the data presented in Fig.~\ref{fig:dU4dl} with the help of very modest numerical efforts (about two weeks of simulations to achieve statistical error below 1\% for $L=48$ in Fig.~\ref{fig:dU4dl} in the domain
$0.55<U_4<0.75$). 

It should also be mentioned that, in general,  in order to find critical indices the data points with smallest statistical errors in the critical domain of $U_B$ should be analyzed. In the cases discussed here and in Ref.\cite{Max} the derivatives of $U_4$ in this domain exhibit extrema. Thus, errors of $U_4$ do not, practically, contribute  to the errors of the derivatives.

\section{Discussion and summary.}\label{sec:dis}

An outstanding problem in the theory of the critical point is the non-analytical contribution to the diameter $n_d=(n_l +n_g)/2$ -- the average of the liquid $n_l$ and gas $n_g$ densities on the approach to the critical point along the coexistence line.  The term $n_d -n_c \sim (-\tau)^{1-\alpha} $ (for $\tau<0$)   appears due to the lack of the symmetry of the Hamiltonian \cite{Pokrovskii} (see also in Ref.\cite{Landau}). [Here $n_c$ is the density at $h=0,\tau=0$]. In 2D $\alpha=0$ and the contribution $\sim \tau \ln(-\tau)$ would be practically impossible to determine following the standard procedure (based on the extrapolation and multi- parametric fits ). The situation in 3D is not much better because $\alpha \sim 0.1$. In Refs. \cite{Fisher2000,Fisher2001,Fisher2003} it was argued that a much stronger non-analyticity $n_d-n_c\sim \tau^{2\beta}$ should be present.  
Determining such a term still presents a significant challenge for the standard approach in 2D due to the small value of the exponent $2\beta =0.25$. The situation in 3D, where $2\beta \approx 0.65$, becomes more optimistic (or, rather, less pessimistic). However, achieving a controlled accuracy based on the standard procedure  is still challenging.      

The NF method allows overcoming the obstacles of the standard approach. Once the angle $\phi_V$ of the coexistence line in the plane $(\epsilon,\tilde{\mu})$, Fig.~\ref{fig:polar}, is found, the value of $n_d$ can be directly obtained by measuring density $n$ along the line and plotting it vs $U_4$ -- in order to guarantee that the system remains in the critical domain for each system size $L$. Then, within the FSS \cite{FSS} 
\beq
n_d = n_c + A L^{-(1-\alpha)/\nu} + B L^{-2\beta/\nu}, 
\label{diam}
\eeq   
where $A,B$ are some non-universal coefficients. 

It  is also possible to measure directly $dn/d\tau \sim dn/dl$. [This quantity can be expressed in terms of
the cumulants $\langle N^2\rangle,\, \langle NN_p\rangle$ and the angle $\phi_V$].
 Then, in the critical domain, $ dn/dl \sim L^{\alpha/\nu}$, if $B=0$, and $ dn/dl \sim L^{(1 -2\beta)/\nu}$, if $B\neq 0$ for large enough $L$. The term predicted in Refs. \cite{Fisher2000,Fisher2001,Fisher2003} will correspond
to the divergence $dn_d/dl \sim L^{0.75}$ in 2D and $\sim L^{0.55}$ in 3D.  This project will be discussed elsewhere.

It is worth mentioning that the flowgram method can be used for finding and analyzing the multicritical points \cite{Annals}. It is based on observing a separatrix in the flow of a properly chosen Binder cumulant $U_B$ --  toward a value $U_c$ different from those observed away from a tentative multicritical point. Thus, similarly to how it was demonstrated above for the LG critical point, plotting the derivatives of $U_B$ vs $U_B$ in the domain around the asymptotic separatrix value $U_c$ would allow extracting the critical exponents.   

In summary, the numerical flowgram method \cite{Annals} has been further developed and demonstrated to be an effective tool in the FSS study of the LG critical point in 2D. Its main advantage is in avoiding the normally used procedures where the position of the critical point and the coexistence line are treated as the fitting parameters in the extrapolation procedure. The only fitting procedure used here is for finding the critical indices $\mu, \nu$ from the FSS relation between the derivatives of the Binder cumulant. These have been determined to be consistent with the Onsager values within 1-2\% of the combined error.  The position of the coexistence line is found with much better accuracy  -- up to 0.1-0.2\% of the total error. This opens up a possibility for resolving the non-analytical corrections to the diameter with the controlled accuracy.

\section { Acknowledgments}
This work 
was supported by the National Science Foundation under the Grant DMR1720251. Simulations have been supported under the Grants CNS-0958379, CNS-0855217, ACI-1126113 and the City University of New York High Performance Computing Center at the College of Staten Island.

\end{document}